\documentclass[12pt, draftclsnofoot, onecolumn]{IEEEtran}
\IEEEoverridecommandlockouts
\usepackage{cite}
\usepackage{amsmath,amssymb,amsfonts}
\usepackage{algorithmic}
\usepackage{graphicx}
\usepackage{textcomp}
\usepackage{xcolor}
\usepackage{verbatim}
\usepackage{mathtools, nccmath}

\usepackage{multirow}
\usepackage{array}

\ifCLASSOPTIONcompsoc
    \usepackage[caption=false, font=normalsize, labelfont=sf, textfont=sf]{subfig}
\else
\usepackage[caption=false, font=footnotesize]{subfig}

\def\BibTeX{{\rm B\kern-.05em{\sc i\kern-.025em b}\kern-.08em
    T\kern-.1667em\lower.7ex\hbox{E}\kern-.125emX}}
\begin{document}
This work has been submitted to the IEEE for possible publication. Copyright may be transferred without notice, after which this version may no longer be accessible.
\newpage
\title{Frame based equipment channel access enhancements in NR unlicensed spectrum for the URLLC transmissions\\
}

\author{\IEEEauthorblockN{Trung-Kien Le$^{\dagger}$, Umer Salim
, Florian Kaltenberger$^{\dagger}$}\\
\IEEEauthorblockA{$^{\dagger}$EURECOM, Sophia-Antipolis, France \\
Trung-Kien.Le@eurecom.fr}
}

\maketitle

\begin{abstract}
  \normalsize
Ultra-reliable low-latency communication (URLLC) in 5G New Radio has been originally defined only for licensed spectrum. However, due to new use cases in the Industry 4.0 scenarios, URLLC operation is currently being extended to unlicensed spectrum in the ongoing Release 17 of the 3rd Generation Partnership Project. Although in such controlled environments we can guarantee the absence of any other technology sharing the channel on a long-term basis, the uncertainty of obtaining channel access through load based equipment (LBE) or frame based equipment (FBE) can impede with the latency requirements of URLLC.
In FBE, the transmitters can be prioritized to support data with different requirements and have lower energy consumption and latency compared to LBE with a big contention window size. In this paper we analyze the performance of FBE in an unlicensed controlled environment through a Markov chain. Based on this analysis, we propose two schemes to improve the URLLC performance in FBE: The first scheme allows the transmitters to use multiple fixed frame period (FFP) configurations while the second scheme configures the FFP's starting point of each transmitter based on its priority. The simulations show the benefits of these schemes compared to the URLLC transmission of existing schemes.

\end{abstract}


\section{Introduction} \label{I}

\subsection{Ultra-reliable low-latency communications (URLLC)} \label{IA}

Many emerging applications such as autonomous driving, Industry 4.0 and augmented reality require the communications with high reliability and low latency. To satisfy these new requirements, the 3rd Generation Partnership Project (3GPP) included URLLC as one of three service categories in 5G new radio (NR). URLLC has attracted the attention on physical layer design since 3GPP Release 15 that is the first full set of 5G standard due to its strict latency and reliability requirements compared to Long-Term Evolution (LTE). 

To satisfy the requirements of the emerging applications and set a baseline for URLLC design, the URLLC requirements are defined in \cite{ref1}: ``A general URLLC reliability requirement for one transmission of a packet is 10\textsuperscript{-5} for 32 bytes with a user plane latency of 1 ms''. These requirements pose a challenge to URLLC design because URLLC reliability of 10\textsuperscript{-5} is much higher than block error rate of 10\textsuperscript{-2} in LTE. URLLC design even becomes more challenging in Release 16 when higher requirements are specified for new use cases such as factory automation, transport industry and electrical power distribution: ``Higher reliability (up to 10\textsuperscript{-6}), higher availability, short latency in the order of 0.5 to 1 ms, depending on the use cases (factory automation, transport industry and electrical power distribution)''\cite{ref2}. These requirements demand that new features must be standardized to support URLLC in 5G.

\subsection{URLLC physical layer's features in 3GPP Release 15 and Release 16} \label{IB}

Several URLLC features have been specified in 3GPP Release 15 - the first full set of 5G standard and Release 16 - the latest finalized release - to improve NR system performance based on URLLC requirements.

The values of subcarrier spacing in 5G NR are 15, 30, 60, 120 and 240 kHz instead of a single value of 15 kHz in LTE. This feature reduces symbol length that allows a shorter transmission in time \cite{ref5}.

In LTE, a transmission time interval (TTI) is one slot and a transmission only can start at the fixed symbols at the beginning of a slot. If a packet arrives after the starting symbols in a slot, it must wait until the next slot to be transmitted. This delay is harmful to URLLC with a low latency requirement. To reduce the waiting time of a packet, in 5G NR, a TTI can be a sub-slot of 2, 4 or 7 symbols and a transmission can start at the beginning of a sub-slot \cite{ref5}.

In NR uplink (UL) transmission, the user equipment (3GPP terminology: UE) is allowed to transmit data to the base station (3GPP terminology: gNB) in the configured grant (CG) resources without scheduling request and UL grant to reduce latency. A UE might have several configurations of the CG resources to choose to start a transmission based on the arrival time of a packet. To increase UL transmission's reliability, the UE might be configured to transmit the repetitions of a packet in the consecutive slots or sub-slots without feedback from the gNB \cite{ref6}.

When the UE having high priority traffic such as URLLC coexists with the UE having low priority traffic such as eMBB, two techniques have been standardized to guarantee URLLC UL transmission's performance. In the first technique, the eMBB UE is ordered by a control signal from the gNB to stop its UL transmission overlapping with the URLLC UL transmission of another UE. The second technique allows the URLLC UE to increase power level of its UL transmission overlapping with the UL transmission of an eMMB UE to compensate the impact of interference \cite{ref17}.

\subsection{URLLC in unlicensed spectrum}\label{IC}

The URLLC features standardized in 3GPP Release 15 and Release 16 are for the operation in licensed spectrum. With the new use cases in the industrial scenario, unlicensed spectrum starts to attract the attention because of its low cost, high flexibility, simplicity of deployment and availability of bandwidth. URLLC operation in unlicensed spectrum is one of the main work items of 3GPP standardization starting from the ongoing Release 17. 

The features of New Radio-Unlicensed (NR-U) have been specified since Release 15 \cite{ref3}. However, these features do not take into account the URLLC features and requirements. The latency and reliability requirements of URLLC might not be achieved in unlicensed spectrum due to listen before talk (LBT). In unlicensed spectrum, a transmitter is required to do channel access mechanism through LBT before acquiring a channel in a certain amount of time to transmit data. There are two channel access mechanisms: load based equipment (LBE) and frame based equipment (FBE) \cite{ref15}. In LBE, a transmitter can do channel sensing to obtain a channel at any moment that it has data to transmit. The impact of LBE on URLLC is analyzed in \cite{ref18}. The transmitter using LBE with a small contention window size has low latency in channel access but a small contention window size causes an inter-system unfairness. On the other hand, with a big contention window size, the transmitter has high latency in channel access and low throughput. In FBE, if a transmitter has a packet to transmit, it only can sense the channel and start a transmission at the fixed moments. FBE benefits the URLLC nodes when data rate is low and data arrival is periodic because the receiver does not always have to detect blindly the presence of a transmission at any moments. It reduces the burden and energy consumption at the receiver. Moreover, URLLC UL data is transmitted in the periodical CG resources. If the CG resources is aligned with the starting points of FBE periods, the transmitter is set to only do LBT at the moments with transmission potential and the receiver only has to listen to the channel at those moments. Furthermore, because the transmitters sense channel and transmit data at the fixed moments, the transmitters can be prioritized based on data requirements by setting the parameters relalating to these moments such as periodicity or starting point.

In this work, the performance of FBE channel access mechanism is analyzed to see its impact on URLLC operation in unlicensed spectrum. The targeted scenario is the industrial scenario in the factories where the absence of any other technologies sharing the channel such as WIFI and devices operating in LBE is guaranteed on a long-term basis. It can be done by the facility owner when he installs the devices to prevent an unexpected interference from other systems and radio access technologies. This environment is called an unlicensed controlled  environment \cite{ref35}, \cite{ref36}. The industrial scenario with an unlicensed controlled  environment is chosen to follow the work of the ongoing 3GPP Release 17 where one of the objectives is the UL enhancements for URLLC in an unlicensed controlled  environment \cite{ref37}. However, this does not mean that LBT is not required prior to a transmission in an unlicensed controlled environment. Although the URLLC nodes work in a controlled environment without WIFI and devices operating in LBE, the uncertainty of FBE LBT due to a competition among the nodes still makes the URLLC nodes unable to attain the requirements. To further improve FBE performance for URLLC, new schemes will be studied.

\subsection{Related work }\label{ID}

\cite{ref4} shows the throughput of a system using FBE on 5 GHz unlicensed band. \cite{ref7} and \cite{ref10} show the performance of FBE when LTE and WIFI coexist. These models in \cite{ref4}, \cite{ref7} and \cite{ref10} do not include the constraints of the URLLC transmission so no enhancements of FBE for URLLC operation are considered. In \cite{ref19}, the model to analyze the coexistence of LTE using FBE and WIFI is only for DL transmission from one base station to several UEs in unlicensed spectrum while UL data is transmitted in licensed spectrum.

\cite{ref8} proposes ``Enhanced FBE'' and ``Backoff and Idle Time Reduction FBE''. In Enhanced FBE, a backoff procedure used after a clear channel access (CCA) increases delay in the URLLC transmission. In Backoff and Idle Time Reduction FBE, the idle period is eliminated after an unsuccessful CCA so the transmitters sense the channel after the channel occupancy time in the next frame instead of waiting the entire frame. If the channel occupancy time is long, it still limits the number of channel sensing at the transmitter. Moreover, a backoff procedure is also used that increases latency.

\cite{ref9} proposes that a transmitter senses periodically or continuously the channel in a subframe. The idle period is removed in this subframe compared to the original FBE frame. This method changes the design of the conventional frame that might make it incompatible in the system that the transmitters use both the conventional frame and the proposed frame. Moreover, to increase the sensing opportunities in the URLLC latency budget, the duration of subframe must be reduced that affects the duration of a transmission.

In \cite{ref11} and \cite{ref12}, a transmitter might not transmit after a successful CCA if channel quality is bad to avoid a higher required transmission power. However, when a transmitter does not transmit after a successful CCA, it misses its transmission opportunity and without transmitting data, it has no chance to have a successful transmission of a packet in the URLLC latency budget. Therefore, the latency constraint of URLLC limits the use of this scheme.

In \cite{ref13}, a transmitter does channel sensing on several parallel channels and can switch among the channels to avoid the interference and the busy channels. The transmitter also can change the idle period in a frame when there is no available channel. This scheme consumes more resources for one transmission because the transmitter is scheduled with multiple resources in the parallel channels for a transmission instead of only one resource in a channel in the conventional scheme.

In \cite{ref16}, a central entity is used in a fully coordinated FBE approach to configure a common Time Division Duplex (TDD) configuration among the nodes in the system so that the UE's UL transmission to a gNB is not blocked by the neighbor gNB DL transmission due to the misalignment of UL and DL slots among the gNBs. A common TDD configuration among the gNB nodes might not satisfy the specific requirements of each gNB network about the ratio of DL and UL transmissions.

In summary, the shortcoming of the existing FBE analysis is that they have not taken into account the URLLC requirements so the limitations of FBE affecting the URLLC transmission's performance have not been analyzed and solved. Some schemes are proposed to add a backoff procedure or avoid transmitting due to a bad channel after a successful CCA. These proposals increase transmission latency that is harmful to URLLC. Two works propose to remove the idle period in the frame with some conditions. This allows the device to have more opportunities to sense a channel in the same interval. However, the number of channel sensing in an interval depends on channel occupancy time so it cannot be modified flexibly based on the requirements of a transmission. One scheme uses several parallel channels to transmit data that increases resource consumption when multiple resources are scheduled in several channels for the transmission of a packet.

\subsection{Main contributions}\label{IE}

This paper focuses on the URLLC operation in an unlicensed controlled  environment where the transmitters use FBE to access to a channel. The analysis of URLLC operation's channel access in FBE is done through a Markov chain in Section \ref{II}. Firstly, a relationship among the probability of sensing a busy channel (blocking probability), data rate, the allowed number of channel sensing and the number of the transmitters is derived for all types of data (not only URLLC data).  Subsequently, the analysis shows that for URLLC data, the URLLC latency constraint limits the number of channel sensing and causes an increase of channel access failure's probability. Ultimately, it results in a degradation of the URLLC transmission's reliability.

Based on the analysis, we propose two schemes in Section \ref{III} and Section \ref{IV} to reduce the channel access failure's probability in FBE for the URLLC transmitters so that they can achieve the URLLC reliability and latency requirements. In Section \ref{III}, the multiple FFP configuration scheme is introduced. In this scheme, a transmitter with high priority is configured with multiple FFP configurations in FBE. These FFP configurations are in the same frequency resources but overlap in time so they do not require the additional resources. The transmitter can use these FFP configurations at the same time and switch between the configurations to do channel sensing and transmit data after obtaining a channel. Thanks to the use of multiple FFP configurations, the transmitter has more opportunities to do channel sensing in the URLLC latency constraint than the conventional scheme. Therefore, the probability of channel access failure decreases.

In Section \ref{IV}, a scheme of arranging FFP of each transmitter based on its priority in the system is presented. The high priority transmitters such as the URLLC transmitters are set up with the offsets for the starting points of FFP so that their transmissions are not blocked by the transmissions of the lower priority transmitters. Thereby, the high priority transmitters achieve the lower channel access failure's probabilities.

Section \ref{V} shows the simulations to verify the analysis in the previous sections. The simulations are also done to compare the conventional scheme and the schemes in the references with two proposed schemes. Based on the results, some suggestions about which proposed scheme is used in a specific scenario are made. Finally, Section \ref{VI} concludes this work.

\section{Analysis of FBE in unlicensed spectrum}\label{II}

\subsection {System model}\label{IIA}
\begin{figure}[htbp]
\centerline{\includegraphics[scale=0.35]{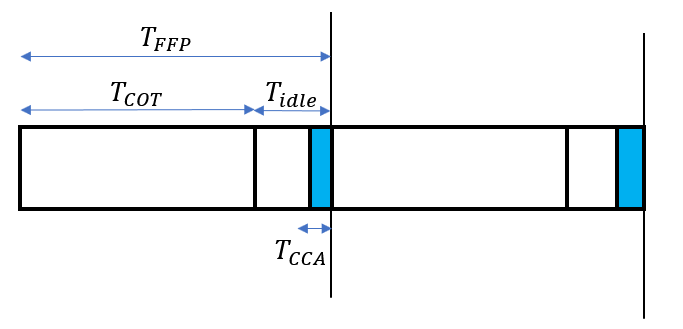}}
\caption{Fixed frame period in FBE.}
\label{fig1}
\vspace{-3mm}
\end{figure}

\begin{figure*}[htbp]
\centerline{\includegraphics[scale=0.57]{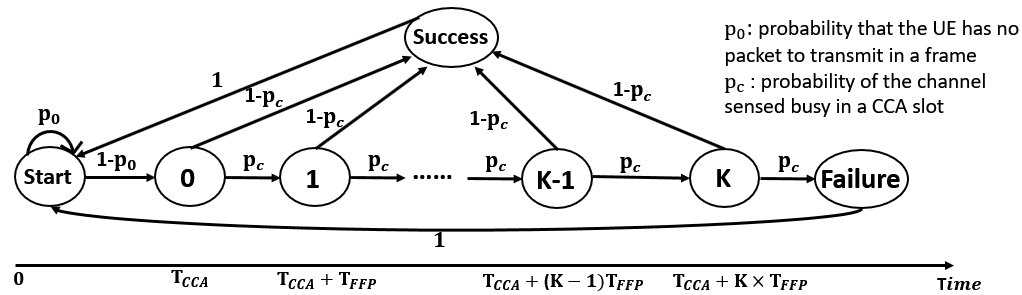}}
\caption{The Markov chain for FBE channel access.}
\label{fig2}
\vspace{-3mm}
\end{figure*}

In FBE channel access mechanism, when a transmitter has data to transmit, it senses a channel to check its availability for a transmission only per fixed period called fixed frame period (FFP) with a duration of 1, 2, 2.5, 4, 5 or 10 ms. As shown in Fig.~\ref{fig1}, a FFP consists of a channel occupation time (COT) and an idle period with the durations of $T_{COT}$ and $T_{idle}$, respectively. The maximum duration of $T_{COT}$ is 95\% of $T_{FFP}$. The duration of $T_{idle}$ is at least 5\% of $T_{FFP}$ but not smaller than 100 $\mu$s. In the idle period, there is a single observation slot where a CCA within $T_{CCA}$ of 25 $\mu$s  is performed. If a transmitter has data and senses an idle channel at a CCA occasion, the transmitter starts immediately a transmission at the beginning of a FFP after the end of that CCA occasion. The transmitter occupies the channel in $T_{COT}$ and stops the transmission in $T_{idle}$. The transmitter can share the channel within $T_{COT}$ with the receiver so that the receiver is also able to use that COT to transmit data. In contrast, if the transmitter senses a busy channel, it does not transmit data and has to wait until the CCA occasion in the next FFP to perform another channel sensing. In a network, both the gNB and the UE are capable of initiating its own COT.

FBE channel access mechanism is modeled by a Markov chain in Fig.~\ref{fig2}. $p_0$ is the probability that a transmitter has no packet to transmit in a FFP. If the transmitter has a packet to transmit in a FFP, it jumps from State Start to State 0 with a probability of $1-p_0$ and senses the channel in a CCA occasion at the end of that frame. Ignoring the alignment time, time consumption for the first channel sensing is $T_{CCA}$. If the channel is sensed idle, the transmitter obtains the channel to transmit at $T_{CCA}$ and jumps to State Success with a probability of $1-p_c$ then goes back to State Start where $p_c$ is the probability of sensing a busy channel. State Success means that the transmitter succeeds in acquiring the channel. It does not means that after acquiring the channel, the transmitter transmits data and data is decoded correctly at the receiver.  If the channel is busy, the transmitter must wait until the CCA occasion in the next FFP to continue to sense the channel and jumps from State 0 to State 1 with a probability of $p_c$. If the sensing is successful, the transmitter obtains the channel at $T_{CCA}+T_{FFP}$. The process continues until the transmitter accesses to the channel or gives up the sensing process for that packet due to a time constraint. In the model, State K presents the last channel sensing at the (K+1)th frame since the first sensing frame. K can be infinite. 

The Markov chain in Section \ref{II} is used to model FBE channel access mechanism of the URLLC transmitters or other types of the 5G NR transmitters coexisting with the URLLC transmitters in the 5G NR system. The transmitters share the same frequency channel in sub-6GHz bands, use omnidirectional sensing (omni-LBT) to sense and acquire a channel in FBE channel access mechanism then transmit data by using omnidirectional transmission. Every of the transmitters can detect each other. The hidden nodes are not included because they do not affect the ability of a transmitter to access to a channel that is the main focus of this Markov chain model. The transmitter cannot sense the hidden nodes so it is not blocked to access to the channel by these hidden nodes. The receivers use omnidirectional reception to receive data. The URLLC transmitters have the transmissions with the strict requirements of latency and reliability so they work in an unlicensed controlled environment to satisfy these requirements because the unlicensed controlled environment guarantees the absence of other access technologies such as WIFI and the devices in 5G network operating in LBE on a long-term basis \cite{ref35}, \cite{ref36}. The unlicensed controlled  environment is also applied in the scenarios in Section \ref{III} and Section \ref{IV}. 

\subsection{Probabilities of the states and channel access in Markov chain for FBE channel access} \label{IIB}

We denote $\pi_{Start}, \pi_{Success}, \pi_{Failure}, \pi_i$ where $i \in [0, K]$ to be stationary distributions of Markov chain in Fig.~\ref{fig2}. We have 
\begin{equation}
\pi_{Start}=\pi_{Success}+\pi_{Failure}.\label{eq1}
\end{equation}

The probability of State 0 is

\begin{equation}
     \pi_{0} = (1-p_0)\pi_{Start}. \label{eq2}
\end{equation}

The probabilities from States 1 to States K are: 

\begin{equation}
     \pi_i = p_c^i\times\pi_0 \hspace{1cm} i \in [1, K]. \label{eq3}
\end{equation}

The probability that a transmitter succeeds in obtaining a channel is also the probability that a transmitter sends data. The probability of a channel sensing success $\pi_{Success}$ is

\begin{align}
    \pi_{Success}&=\pi_0(1-p_c)\sum_{i=0}^{K}p_c^i  \nonumber\\
    &=\pi_0(1-p_c^{K+1})\nonumber\\
    &=\pi_{Start}(1-p_0)(1-p_c^{K+1}).\label{eq5}
\end{align}

The probability is calculated when a packet already entered the process so $\pi_{Start}$ equals to 1. In other words, the transmission probability is calculated given that $\pi_{Start}$ is 1. By substituting $\pi_{Start}=1$ to \eqref{eq5}, we have the probability that a transmitter transmits data

\begin{equation}
    P_{trans}=(1-p_0)(1-p_c^{K+1}).\label{eq6}
\end{equation}

When a transmitter has a packet and senses the channel to transmit that packet, the probability that the transmitter fails to obtain the channel after all allowed channel sensing and must drop the packet is

\begin{equation}
    P_{failure}=p_c^{K+1}.\label{eq7}
\end{equation}

\subsection{Relation between $p_0$ and $p_c$} \label{IIC}

\begin{figure}[htbp]
\centerline{\includegraphics[scale=0.3]{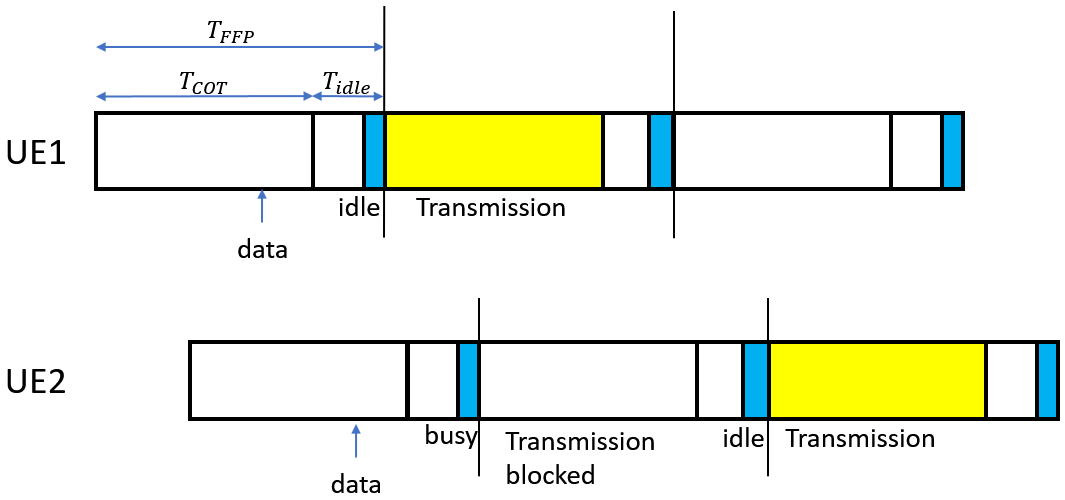}}
\caption{Channel sensing and data transmission in FBE.}
\label{fig5}
\vspace{-3mm}
\end{figure}

The FFP of each transmitter can be configured with an offset to start at different time in the same frequency resources as Fig.~\ref{fig5} so a CCA occasion of a transmitter might overlap with a COT of another transmitter. If a transmitter is sending data in a COT then another transmitter senses the channel at that time, the channel is sensed busy as the case of UE2 in Fig.~\ref{fig5}. The UE2 must wait the next FFP to sense the idle channel and transmit data. This offset ensures two UE not to sense an idle channel then transmit at the same time so as to avoid the interference between two simultaneous transmissions causing a degradation of transmission's reliability.

We consider a system that has $Q$ transmitters with the same $K+1$ sensing states but different starting points of FFPs. These transmitters belong to a 5G NR network and send data in the same frequency resource in an unlicensed controlled  environment. A CCA occasion of a transmitter overlaps with the COTs of the other transmitters. A transmitter of interest senses an idle channel in a CCA occasion if there is no UE out of $Q-1$ UE transmitting at that time. We have the relation between $p_0$ and $p_c$ by using the transmission probability of a transmitter in \eqref{eq6}

\begin{equation}
    p_c=1-(1-(1-p_0)(1-p_c^{K+1}))^{Q-1}.\label{eq8}
\end{equation}

From \eqref{eq8}, $p_c$ is calculated when $p_0, K, Q$ are known.

\subsection{URLLC operation with FBE in unlicensed spectrum} \label{IID}

As shown in Fig.~\ref{fig5}, a transmitter does channel sensing to transmit a packet in the fixed moments with the gap of $T_{FFP}$ between two consecutive moments. The URLLC transmission has a latency budget of 1 ms while the smallest duration of a fixed frame period $T_{FFP}$ is 1 ms. Therefore, in the URLLC latency budget, the transmitter only can do one channel sensing because the second channel sensing is at $T_{CCA}+T_{FFP}$ that is bigger than 1 ms. If it fails in this only chance, it cannot access to the channel to transmit data in the URLLC latency requirement and the URLLC packet is dropped.

When a transmitter has a URLLC packet with only one chance of channel sensing, the value of $K$ in the Markov chain is 0. Substituting $K=0$ to \eqref{eq7}, we have the probability that the transmitter fails to access to a channel in order to transmit an URLLC packet is:

\begin{equation}
    P_{failure\_URLLC}=p_c.\label{eq9}
\end{equation}

When $Q$ transmitters are configured to transmit the URLLC packets in a controlled environment, from \eqref{eq8}, we have the relation between $p_0$ and $p_c$ of the URLLC transmitters.

\begin{equation}
    p_c=1-(1-(1-p_0)(1-p_c))^{Q-1}.\label{eq10}
\end{equation}

The limit of channel sensing opportunity due to the time constraint increases the probability of channel access failure. It causes an increase of the dropped packets and reduces the URLLC transmission's reliability.

This section provided a Markov chain to model channel access in unlicensed spectrum when the transmitters in the system use FBE to acquire a channel. Subsequently, the URLLC latency constraint is applied to the model to show the impact of channel access in FBE on URLLC performance.

\section{Multiple configurations of FFP in FBE for URLLC in unlicensed spectrum}\label{III}

\subsection{Multiple configurations of FFP}\label{IIIA}

\begin{figure}[htbp]
\centerline{\includegraphics[scale=0.34]{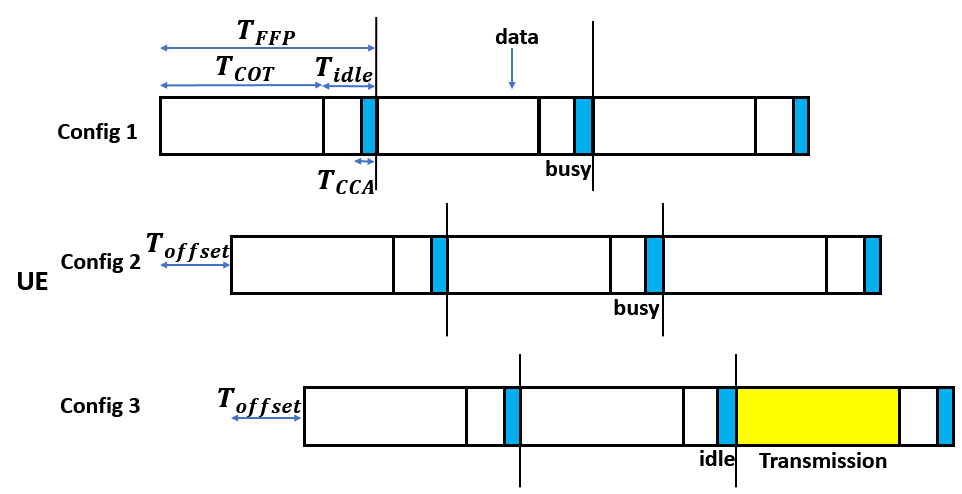}}
\caption{Multiple configurations of FFP.}
\label{fig3}
\end{figure}

The conventional FBE scheme provides only one opportunity of channel sensing in the URLLC transmission due to the time constraint. To provide more opportunities of channel access in the URLLC latency budget of 1 ms and reduce the errors due to the packets dropped out of the latency budget in UL transmission, we propose a new FBE scheme where a transmitter is configured with multiple configurations of FFP. These FFP configurations are in the same frequency resources but overlap in time and have different starting points so they do not require the additional resources compared to the conventional scheme with one FFP configuration as shown in Fig.~\ref{fig3}. The number of FFP configurations that a transmitter can use is configured by the gNB through downlink control information (DCI) or radio resource control (RRC) based on arrival rate of data, the probability of channel access, the priority and requirements of data transmission. The beginning of a FFP in a configuration is shifted an amount of time defined as $T_{offset}$ compared to the beginning of a FFP in the previous configuration. $T_{offset}$ is configured by gNB through DCI or RRC based on the number of configurations that a transmitter has.

The transmitter with high priority data such as URLLC uses multiple FFP configurations in the attempts to access to the channel and transmit a high priority packet while the transmitter with low priority data uses only one FFP configuration in the attempts to access to the channel and transmit a low priority packet. This means that a transmitter might have multiple FFP configurations but only uses one configuration in the attempts to access to the channel and transmit low priority data as the conventional FBE scheme. FFP periodicity in all configurations of a transmitter is the same and configured by the gNB through DCI or RRC.

When a transmitter has high priority data such as URLLC, it senses a channel in the CCA occasions of different FFP configurations. It starts to sense the channel from the configuration with the closest CCA occasion from the arrival time of data in order to reduce the waiting time. If it fails to access to the channel in a configuration, it does another attempt in the closest CCA occasion of the next configuration instead of waiting one frame period in the same configuration as the conventional scheme. Subsequently, if it succeeds in channel access in a configuration, it uses that configuration to start the transmission in $T_{COT}$ of a FFP immediately after that successful CCA occasion. This COT is also shared with the receiver to transmit data in the opposite direction.

As shown in Fig.~\ref{fig3}, a UE is configured with three FFP configurations having different starting points in the same frequency resources. When data arrives, the UE starts to do channel sensing in a CCA occasion of the first configuration because this configuration has the closest CCA occasion from the moment that data arrives. This reduces the waiting time before the first sensing. The UE senses a busy channel in the CCA occasion of the first configuration then it moves to the second configuration to do channel sensing in the next CCA occasion. Consequently, the channel is still busy so the UE moves to the third configuration to do channel sensing. This time the channel is idle so the UE chooses this configuration to transmit data. Data is transmitted immediately in a FFP after the successful CCA. With the multiple FFP configuration scheme in this example, the UE has three opportunities of channel sensing in an interval of $T_{FFP}$ instead of only one opportunity in the conventional scheme.

\subsection{The Markov chain of FBE channel access with multiple configurations of FFP}\label{IIIB}

\begin{figure*}[htbp]
\centerline{\includegraphics[scale=0.57]{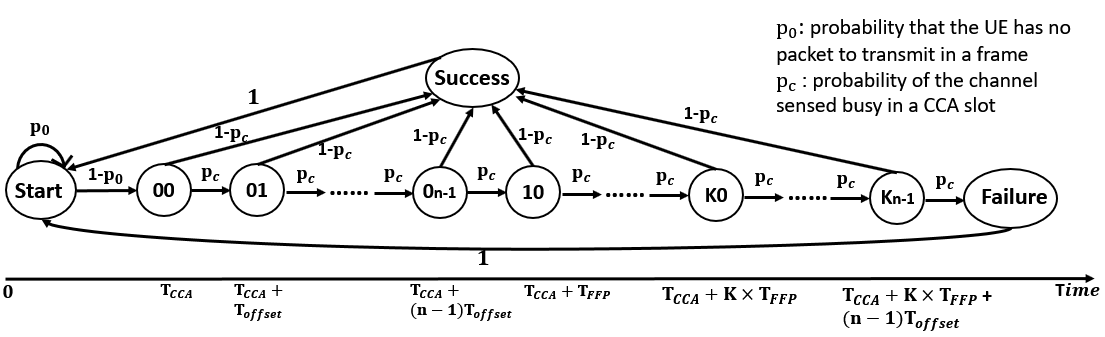}}
\caption{The Markov chain for multiple configurations of FFP in FBE channel access.}
\label{fig4}
\vspace{-4mm}
\end{figure*}

Fig.~\ref{fig4} shows the Markov chain of FBE channel access with multiple FFP configurations. $n$ is the number of the FFP configurations that a transmitter uses to do channel access for a packet's transmission. $T_{offset}$ equals to $\frac{T_{FFP}}{n}$. Other parameters are defined in Section \ref{IIA}. A transmitter has a packet and jumps from State Start to State 00 with a probability of $1-p_0$. State 00 presents the first sensing frame in the first chosen configuration. If the channel is idle, the transmitter jumps from State 00 to State Success with a probability of $1-p_c$ then moves back to State Start. State Success means that the transmitter succeeds in acquiring the channel. It does not means that after acquiring the channel, the transmitter transmits data and data is decoded correctly at the receiver. If the channel is busy, the transmitter jumps from State 00 to State 01 with a probability of $p_c$. State 01 presents the first sensing frame in the next configuration. If the channel continues to be busy, the transmitter goes to the next states. After going through the first sensing frame of all configurations, the transmitter comes back to the first chosen configuration and senses the channel in the second sensing frame as presented by State 10. The process continues until the channel is obtained successfully and the transmitter jumps to State Success or all CCAs in the allowed time fail and the transmitter jumps to State Failure. Finally, the transmitter comes back to State Start.

Following the same calculations in Section \ref{IIB}, when $\pi_{Start}$ equals to 1 for a packet that already entered the process, the probability of transmission for a transmitter using FBE with multiple FFP configurations is 

\begin{equation}
    P_{trans\_nconfig}=(1-p_0)(1-p_c^{n(K+1)}).\label{eq11}
\end{equation}

The probability of channel access failure when the transmitter has a packet to transmit is

\begin{equation}
    P_{failure\_nconfig}=p_c^{n(K+1)}.\label{eq12}
\end{equation}

If $Q$ transmitters have $n$ configurations in a controlled environment, the relation between $p_c$ and $p_0$ is

\begin{equation}
    p_c=1-(1-(1-p_0)(1-p_c^{n(K+1)}))^{Q-1}.\label{eq13}
\end{equation}

From \eqref{eq13}, $p_c$ is calculated when $p_0, K, n, Q$ are known.

In the multiple FFP configuration scheme, a transmitter has maximum $n$ attempts to access to a channel from $T_{CCA}$ to $T_{CCA}+T_{FFP}$ (channel sensing at $T_{CCA}+T_{FFP}$ is not counted). Therefore, a transmitter with an URLLC packet has maximum $n$ opportunities of channel sensing in the URLLC latency budget of 1 ms with $T_{FFP}$ of 1 ms instead of only one opportunity in the conventional FBE scheme. We have $m$ ($m\leq n$) to be the number of FFP configurations that the transmitter can use in the URLLC latency budget of 1 ms. The probabilities of transmission and channel sensing failure of a URLLC transmitter are

\begin{equation}
    P_{trans\_nconfig\_URLLC}=(1-p_0)(1-p_c^m).\label{eq14}
\end{equation}

\begin{equation}
    P_{failure\_nconfig\_URLLC}=p_c^m.\label{eq15}
\end{equation}

The relation between  $p_c$ and $p_0$ for $Q$ URLLC transmitters that can use $m$ configurations in the URLLC latency budget of 1 ms in a controlled environment is

\begin{equation}
    p_c=1-(1-(1-p_0)(1-p_c^m))^{Q-1}.\label{eq16}
\end{equation}

Multiple FFP configurations in FBE mitigate channel access failure in the URLLC transmission due to the time constraint. Moreover, the multiple FFP configuration scheme in FBE also reduces the alignment time compared to the conventional FBE with one configuration. When a packet arrives, a transmitter must wait until the closest CCA occasion to do channel sensing. The alignment time is uniformly distributed among two consecutive CCA occasions. For the conventional FBE scheme, the alignment time is $\frac{T_{FFP}}{2}$. While the alignment time is $\frac{T_{offset}}{2}$ for the multiple FFP configuration scheme. Because $T_{offset}$ between two configurations is smaller than $T_{FFP}$, the alignment time of the multiple FFP configuration scheme is smaller than that of the conventional scheme. It benefits the URLLC transmission with a strict latency requirement. On the other hand, in the multiple FFP configuration scheme, the receiver must detect blindly the presence of a transmission more than one time in a duration of $T_{FFP}$. The receiver has to check each configuration to see whether there is a transmission or not because of the uncertainty of channel sensing and arrival of data at the transmitter. The number of blind detections in a duration of $T_{FFP}$ is equal to the number of the FFP configurations. Therefore, to reduce the burden and energy consumption at the receiver, only the high priority transmitters such as URLLC type are allowed to use multiple FFP configurations in channel sensing and transmission.

\section{FFP arrangement based on the transmitter's priority}\label{IV}

\begin{figure}[htbp]
\centerline{\includegraphics[scale=0.4]{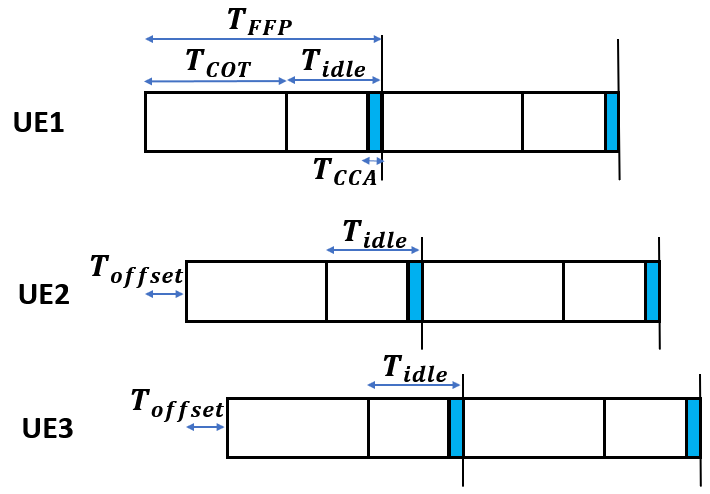}}
\caption{FFP arrangement in FBE based on the UE's priority.}
\label{fig6}
\vspace{-3mm}
\end{figure}

In Section \ref{II} and Section \ref{III}, the transmission in one COT initiated by a transmitter might block any transmitter in the network to initiate its own COT because a CCA occasion of a transmitter overlaps with a COT of another transmitter. It results in the same probability of sensing a busy channel for all transmitters in the network. However, in case there are the transmitters with different priorities such as latency and reliability in a network, it would be better if the high priority transmitters have a higher chance to access to the channel than the low priority transmitters. Therefore, we propose another FBE scheme in this section to support the UE with different priorities. FFP in each transmitter is configured with an offset so that a CCA occasion of a high priority transmitter overlaps with an idle period of a low priority transmitter. In other words, a low priority transmitter stops the transmission before a CCA occasion of a high priority transmitter so the high priority transmitter senses an idle channel to transmit data whenever it has data to transmit. While a CCA occasion of a low priority transmitter overlaps with a COT of a high priority transmitter so the low priority transmitter might be blocked to initiate its own COT. In this scheme, a transmitter is only blocked to initiate its COT by other COT initiators with higher priority and not blocked by other COT initiators with lower priority. The transmitters belong to 5G NR network and send data in the same frequency resources in an unlicensed controlled  environment without interference from WIFI and other devices operating in LBE.

The priorities of the UE transmitters are defined based on the latency and reliability requirements of data that they transmit. Following the 3GPP standards, the priority of data is defined by two methods: a priority indicator field in DCI or RRC. A priority indicator field is added in the new DCI format to indicate the priority of the scheduled data on physical uplink shared channel (PUSCH). However, in CG PUSCH, priority of data on PUSCH is configured by RRC and is not written by the activation DCI.

Fig.~\ref{fig6} shows an example of a FFP arrangement based on the UE's priorities. The UE1 has the highest priority so its CCA occasion is configured to overlap with the idle periods of the UE2 and UE3. The UE1 always senses an idle channel to transmit data and fulfil its requirements. The UE2 has lower priority than the UE1 but higher priority than the UE3. Its CCA occasion is set by $T_{offset}$ to overlap with the UE1's COT and the UE3's idle period. Due to this arrangement, the UE2 might be blocked to initiate its COT by the UE1 but is not affected by the UE3. Finally, the lowest priority UE3 might be blocked to initiate its COT by the other UE because its CCA occasion overlaps with the UE1 and UE2's COT.

The system model with the parameter $p_0$ defined in Section \ref{II} is used to calculate the channel blocking probability of each UE in the FFP arrangement scheme based on the UE's priority. We extend the system in Fig.~\ref{fig6} to a system with $Q$ transmitters. In this system, the UE1 has an absolute priority and is not blocked by the transmissions of the other UE in the network so the probability that the UE1 senses a busy channel is $p_{c1}=0$.

The UE2 has lower priority than the UE1 so it might be blocked by the UE1's transmission. However, it has higher priority than the other UE except the UE1 and is not blocked by those transmissions. The probability that the UE2 senses a busy channel is

\begin{align}
    p_{c2}&=1-(1-(1-p_0)(1-p_{c1}))  \nonumber\\
                &=1-p_0.\label{eq17}
\end{align}

The UE3 might be blocked by the UE1 and UE2's transmissions with higher priorities so the probability that the UE3 senses a busy channel is

\begin{align}
    p_{c3}&=1-(1-(1-p_0)(1-p_{c1}))(1-(1-p_0)(1-p_{c2}))  \nonumber\\
                &=1-p_0(1-(1-p_0)p_0).\label{eq18}
\end{align}

From \eqref{eq17} and \eqref{eq18}, we can derive a general equation of the channel blocking probability for the $i^{th}$ UE in the FFP arrangement scheme based on the UE's priority

\begin{equation}
    p_{ci}=1-\prod_{j=1}^{i-1}(1-(1-p_0)(1-p_{cj})).\label{eq19}
\end{equation}

The URLLC UE has only one chance to do channel sensing so the arrangement to obtain $p_c=0$ for the URLLC UE guarantees the URLLC transmission in the latency budget. Therefore, this scheme benefits the URLLC UE where the URLLC UE coexists with other lower priority UE.

To arrange the CCA occasions of the high priority UE to overlap with the idle periods of the low priority UE, the requirement of the idle period $T_{idle}$ must be stricter than the current requirement of $T_{idle}$ where $T_{idle}$ is at least 5\% of $T_{FFP}$ but not smaller than 100 $\mu$s. $T_{idle}$ in a network with $Q$ UE in the FFP arrangement scheme based on the UE's priority must satisfy

\begin{equation}
    T_{idle} > max\{(Q-1)T_{offset} + T_{CCA}, 100 \mu s\}.\label{eq20}
\end{equation}

$T_{idle}$ increases when the number of the UE in the network increases. When $T_{FFP}$ does not change, it results in a decrease of the transmission time $T_{COT}$. Therefore, the UE cannot transmit a long packet but has to fragment it into small segments and transmits them in the consecutive FFPs. Latency increases because there are more idle periods in the transmission of this long packet. Moreover, the UE needs to do channel access between FFPs that increases latency due to the uncertainty of channel access. Therefore, the arrangement of FFP to guarantee the transmission of the high priority UE is suitable to a network with a small number of the UE having different priorities and short packets to transmit. 

\section{Numerical and simulation results}\label{V}

\begin{table}[htbp]
\normalsize
\caption{Simulation parameters for Fig.~\ref{fig7}, Fig.~\ref{fig8}, Fig.~\ref{fig18}, Fig.~\ref{fig19}, Fig.~\ref{fig22} and Fig.~\ref{fig23} }
\begin{center}
\begin{tabular}{|p{14em}|p{11em}|}
 \hline
 \textbf{Parameters} & \textbf{Values}\\
 \hline
 Fixed frame period & 1 ms\\
 \hline
 Channel occupancy time & 900 $\mu$s\\
  \hline
 Number of configurations per UE & 1-4\\
   \hline
 $p_0$ & 0.99, 0.95\\
  \hline
 Number of simulated frames & $10^{10}$\\
 \hline
 Bandwidth & 20, 80 MHz\\

 
 \hline
\end{tabular}
\label{tab1}
\end{center}
\vspace{-4mm}
\end{table}

In this section, the analytic results are verified by the MATLAB simulations. The performance of channel access in three different schemes of FBE analyzed in Sections \ref{II}, \ref{III} and \ref{IV} are compared: the conventional scheme where each transmitter uses one FFP configuration to sense a channel and transmit data, the proposed scheme where each transmitter uses multiple FFP configurations to sense a channel and transmit data, the proposed scheme where FFPs for the transmitters in the system are arranged based on their priorities.

\begin{figure}[htbp]
\centerline{\includegraphics[scale=0.38]{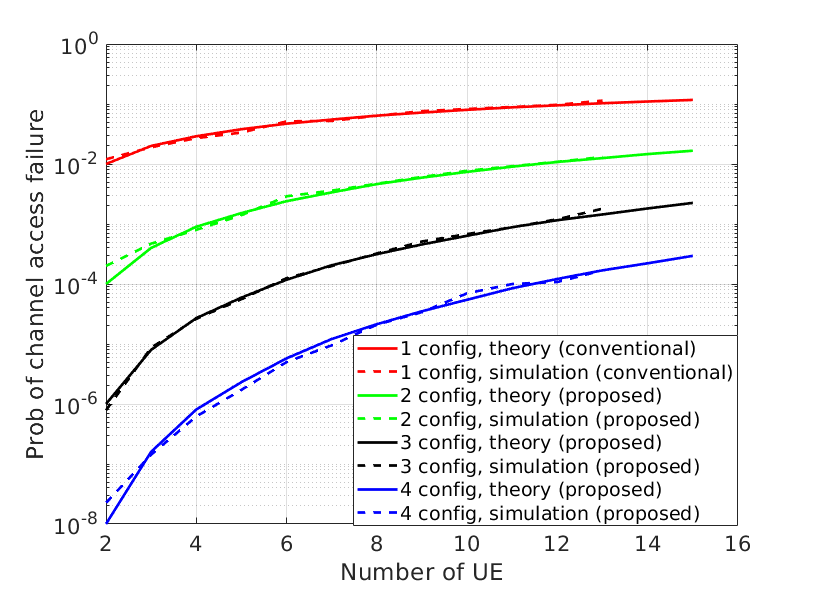}}
\caption{Channel access failure's probability in FBE in the conventional one-configuration scheme and the proposed multiple-configuration scheme with $p_0=0.99$ and bandwidth of 20 MHz.}
\label{fig7}
\vspace{-3mm}
\end{figure}

\begin{figure}[htbp]
\centerline{\includegraphics[scale=0.4]{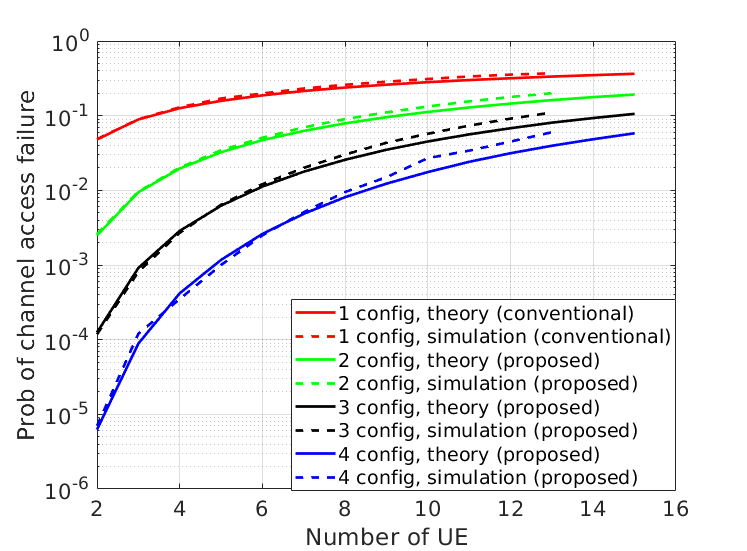}}
\caption{Channel access failure's probability in FBE in the conventional one-configuration scheme and the proposed multiple-configuration scheme with $p_0=0.95$ and bandwidth of 20 MHz.}
\label{fig8}
\vspace{-3mm}
\end{figure}

\begin{figure}[htbp]
\centerline{\includegraphics[scale=0.35]{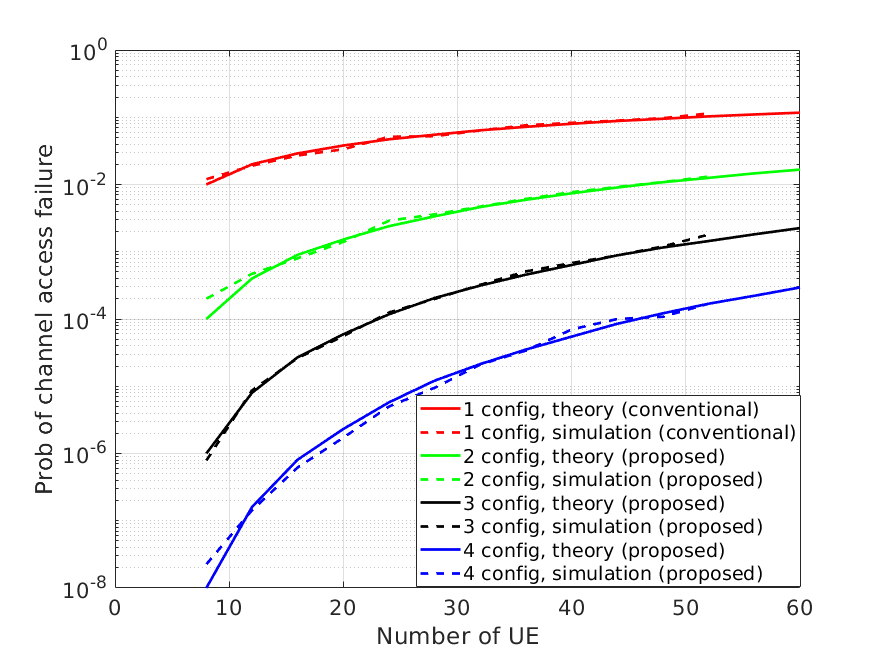}}
\caption{Channel access failure's probability in FBE in the conventional one-configuration scheme and the proposed multiple-configuration scheme with $p_0=0.99$ and bandwidth of 80 MHz.}
\label{fig18}
\vspace{-3mm}
\end{figure}

\begin{figure}[htbp]
\centerline{\includegraphics[scale=0.4]{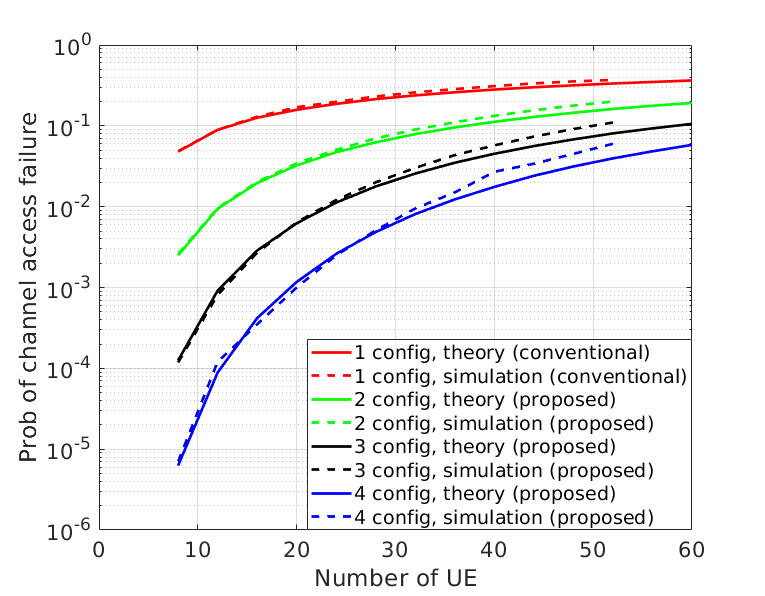}}
\caption{Channel access failure's probability in FBE in the conventional one-configuration scheme and the proposed multiple-configuration scheme with $p_0=0.95$ and bandwidth of 80 MHz.}
\label{fig19}
\vspace{-3mm}
\end{figure}

\begin{figure}[htbp]
\centerline{\includegraphics[scale=0.54]{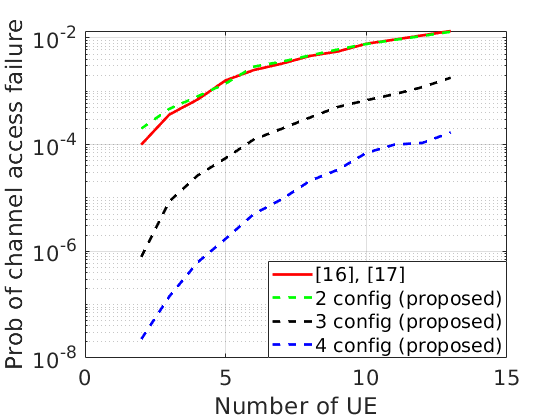}}
\caption{Channel access failure's probability in the schemes of \cite{ref8}, \cite{ref9} and the proposed  multiple-configuration scheme $p_0=0.99$ and bandwidth of 20 MHz.}
\label{fig22}
\vspace{-3mm}
\end{figure}

\begin{figure}[htbp]
\centerline{\includegraphics[scale=0.54]{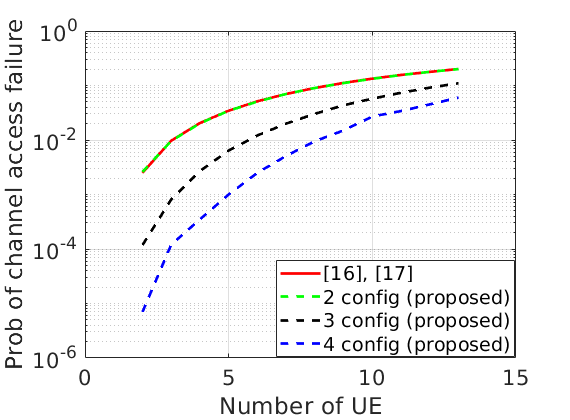}}
\caption{Channel access failure's probability in the schemes of \cite{ref8}, \cite{ref9} and the proposed  multiple-configuration scheme $p_0=0.95$ and bandwidth of 20 MHz.}
\label{fig23}
\vspace{-3mm}
\end{figure}

The parameters in Table~\ref{tab1} are used for the simulations in Fig.~\ref{fig7}, Fig.~\ref{fig8}, Fig.~\ref{fig18}, Fig.~\ref{fig19}, Fig.~\ref{fig22} and Fig.~\ref{fig23}. In these simulations, UL transmission is carried out where the transmitters being the UE transmits data to the receiver being the gNB in a single cell. The transmitters use omnidirectional sensing to sense and acquire a channel in FBE channel access mechanism then transmit data by using omnidirectional transmission. Every of the transmitters can detect each other. The hidden nodes are not included because they do not affect the ability of a transmitter to access to a channel that is the main focus of the proposed scheme. The transmitter cannot sense the hidden nodes so it is not blocked to access to the channel by these hidden nodes. The receivers use omnidirectional reception to receive data.

The FFP is set to 1 ms with COT of 900 $\mu$s. LBT channel access mechanism is done per a bandwidth of 20 MHz so the UE in the simulations uses the same channel with bandwidth of 20 MHz to do channel sensing and transmit data. This means that the number of the UE in the following graphs represents the number of the UE in one frequency resource unit of 20 MHz. This result can be extended to other systems with different bandwidth. In one system, if a bandwidth of 20 MHz is divided into several interlaces and each UE transmits in an interlace, the total number of the UE in the system is the product of the number of the UE shown in  Fig.~\ref{fig7} and Fig.~\ref{fig8} and the number of interlaces in a bandwidth of 20 MHz. On the other hand, if NR-U carrier is wide band that is a multiple of 20 MHz and each UE in the system works in a channel band of 20 MHz, the total number of the UE in the system is a multiple of the number of the UE shown in  Fig.~\ref{fig7} and Fig.~\ref{fig8}. The results of a wide band system with a bandwidth of 80 MHz are shown in Fig.~\ref{fig18} and Fig.~\ref{fig19}. The UE is assumed to transmit the URLLC packets that have a low arrival rate where $p_0$ is set to 0.99 or 0.95.

Fig.~\ref{fig7} and Fig.~\ref{fig8} show the performance of channel access with different number of FFP configurations per UE in a bandwidth of 20 MHz where $p_0$ is set to 0.99 and 0.95, respectively. In Fig.~\ref{fig7}, the conventional scheme where each UE uses only one FPP configuration in channel sensing has a high probability of channel sensing failure. This probability increases rapidly when the number of the UE in the system increases. Even if there are only two UE, the blocking probability of each UE is $10^{-2}$ that is much higher than URLLC reliability requirement. Fig.~\ref{fig8} shows a similar result for one configuration at $p_0$ of 0.95. Therefore, the conventional scheme is not suitable for the URLLC transmission.

Fig.~\ref{fig7} also shows the blocking probability in the multiple FFP configuration scheme at $p_0$ of 0.99 to compare with the blocking probability in the conventional scheme with one FFP configuration. If each UE has two configurations, the blocking probability of $10^{-4}$ for each UE in the system of two UE is much smaller than that of the conventional scheme. When more FFP configurations per UE are used, the UE achieves a smaller blocking probability even if there is a bigger number of the UE in the system. The benefit of multiple FFP configurations is also shown in Fig.~\ref{fig8}, although the probability of channel access failure in Fig.~\ref{fig8} is higher than that in Fig.~\ref{fig7} because of a higher data rate. Therefore, this scheme is suitable for a system where several URLLC UE coexist.

In \cite{ref8}, when a UE senses a busy channel in a CCA occasion of 25 $\mu$s and cannot acquire the channel, the idle period in the following FFP is removed so the UE can sense the channel after the channel occupancy time instead of waiting the entire FFP. Channel occupancy time in Table~\ref{tab1} is 900 $\mu$s. This means that after an unsuccessful CCA, the UE does the second channel sensing after 900 $\mu$s. Therefore, within the URLLC latency budget of 1 ms, an URLLC UE has maximum two sensing opportunities. Similarly, in \cite{ref9}, the idle period does not exist in a frame so the UE also has maximum two channel sensing opportunities in the latency budget of 1 ms. Fig.~\ref{fig22} and Fig.~\ref{fig23} compare the performance of the schemes in \cite{ref8}, \cite{ref9} with the proposed multiple FFP configuration scheme. As can be seen in the figures, the performance of the schemes in \cite{ref8}, \cite{ref9} is equivalent to the proposed scheme with two FFP configurations. The probability of channel access failure in these schemes is still higher than the URLLC requirements. When the number of the FFP configurations increases to three and four, the multiple FFP configuration scheme achieves a better performance with the lower probabilities of channel access failure. The multiple FFP configuration scheme allows the number of the FFP configurations to be modified flexibly based on channel condition and data requirements without affecting the transmission duration in COT while the schemes in \cite{ref8}, \cite{ref9} must reduce COT to increase channel sensing opportunities.

\begin{table}[htbp]
\normalsize
\caption{Simulation parameters for Fig.~\ref{fig15}}
\begin{center}
\begin{tabular}{|p{14em}|p{11em}|}
 \hline
 \textbf{Parameters} & \textbf{Values}\\
 \hline
 Fixed frame period & 1 ms\\
 \hline
 Channel occupancy time & 650 $\mu$s\\
  \hline
 Offset & 40 $\mu$s\\
   \hline
 $p_0$ & 0.99, 0.95\\
  \hline
 Number of simulated frames & $10^{10}$\\
 \hline
 Bandwidth & 20 MHz\\

 
 \hline
\end{tabular}
\label{tab3}
\end{center}
\vspace{-4mm}
\end{table}

\begin{figure}[htbp]
\centerline{\includegraphics[scale=0.54]{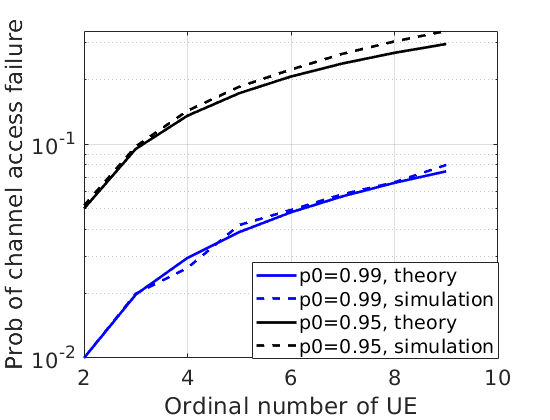}}
\caption{Performance of channel access in the FFP priority arrangement scheme.}
\label{fig15}
\vspace{-3mm}
\end{figure}

The parameters in Table~\ref{tab3} are used for the simulations in Fig.~\ref{fig15}. Fig.~\ref{fig15} shows the performance of channel access of the UE in FBE where FFPs are arranged based on the UE's priorities. The first UE is assumed with the highest priority then the priority of the UE decreases in terms of the ordinal number of the UE. The first UE always has the probability of channel access failure to be 0. In Fig.~\ref{fig15}, each point represents the blocking probability of the $i^{th}$ UE in the system. With $p_0$ of 0.99, the second UE has the blocking probability of 0.01 in the system with at least two UE. The third UE has the blocking probability of 0.02 in the system with at least three UE. Similarly, the blocking probability of the $i^{th}$ UE is presented. The UE is in one bandwidth of 20 MHz and the number of the UE can be extended by using the interlaces or a wide band as explained above.

\begin{figure}[htbp]
\centerline{\includegraphics[scale=0.54]{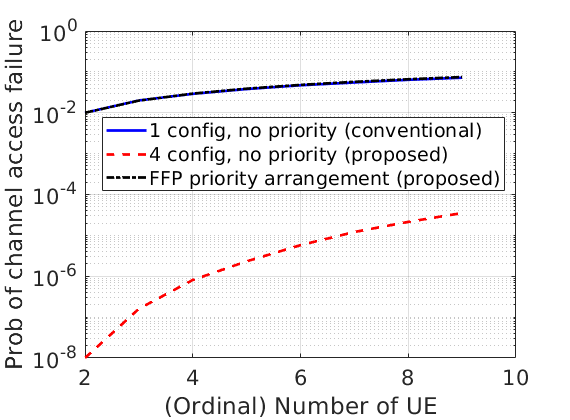}}
\caption{A comparison of the conventional one-configuration scheme, the multiple-configuration scheme and the FFP priority arrangement scheme at $p_0=0.99$.}
\label{fig16}
\vspace{-3mm}
\end{figure}

Fig.~\ref{fig16} compares the performance of channel access in three analyzed schemes. The scheme of FFP priority arrangement gives an approximate probability of channel access failure as the conventional scheme with one FFP configuration. The difference is that all UE in the conventional scheme have the same blocking probability. While the UE in the FFP priority arrangement scheme has the blocking probability depending on its priority. As can be seen in Fig.~\ref{fig16}, for the line of the conventional FBE scheme, each point represents the whole number of the UE in the system. While for the line of the proposed FFP priority arrangement scheme, each point represents the blocking probability of the $i^{th}$ UE (the ordinal number) in the system. For example, if there are three UE, in the conventional scheme, all three UE have the probability of 0.02. While in the FFP priority arrangement scheme, the first UE has the failure probability of 0 that is smaller than the probability of the conventional scheme. The second UE has the probability of 0.01 corresponding to the point (2, 0.01) in the graph that is smaller than the probability of the conventional scheme. The third UE has the probability of 0.02 corresponding to the point (3, 0.02) in the graph. Therefore, the FFP priority scheme is suitable for a system where the UE with different priorities including the URLLC UE coexist. This scheme does not increase the complexity of each UE and network design as the multiple FFP configuration scheme while the URLLC UE is provided an absolute priority at cost of the channel access probability of the other UE. On the other hand, the multiple FFP configuration scheme with four configurations in the simulation provides the best performance of channel access for all UE in the system, although it requires a more complex design of the receivers to detect a transmission in one of four FFP configurations.

\begin{table}[htbp]
\normalsize
\caption{\normalsize Simulation parameters for Fig.~\ref{fig17}}
\begin{center}
\begin{tabular}{|p{14em}|p{11em}|}
 \hline
 \textbf{Parameters} & \textbf{Values}\\
 \hline
 Fixed frame period & 1 ms\\
 \hline
 Channel occupancy time & 900 $\mu$s\\
  \hline
 $p_0$ of URLLC UE & 0.99\\
   \hline
 $p_0$ of low priority UE& 0.5\\
 \hline
Number of URLLC UE & 1-9\\
   \hline
 Number of low priority UE& 1\\
 \hline
 Bandwidth & 20 MHz\\

 
 \hline
\end{tabular}
\label{tab2}
\end{center}
\vspace{-4mm}
\end{table}

\begin{figure}[htbp]
\centerline{\includegraphics[scale=0.54]{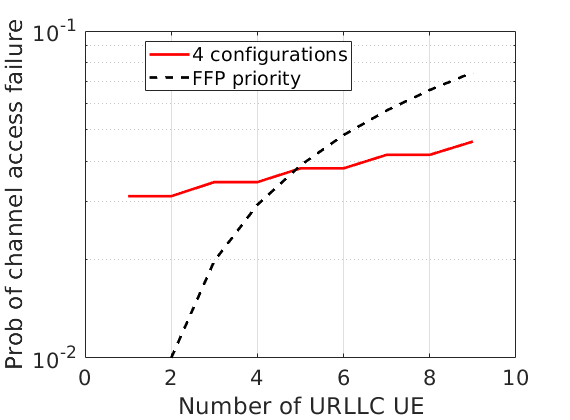}}
\caption{\normalsize A comparison of FBE performance in the multiple-configuration scheme and the FFP priority arrangement scheme in a scenario of the URLLC UE coexisting with the low priority UE.}
\label{fig17}
\vspace{-3mm}
\end{figure}

The parameters in Table~\ref{tab2} are used to simulate a scenario where several URLLC UE coexist with a low priority UE such as an eMBB UE as shown in Fig.~\ref{fig17}. The low priority UE has a higher arrival rate of data and no latency constraint. It can do channel sensing until it obtains the channel so K in the Markov chain goes to infinity for this low priority UE. In the multiple configuration scheme, each URLLC UE has four configurations to sense a channel while the eMBB UE only uses one configuration for channel sensing. In the FFP priority scheme, the eMBB UE is set to the lowest priority FFP. The FFP priority scheme brings the high priority UE a better channel sensing performance. From the first URLLC UE to the fifth URLLC UE in the FFP priority scheme achieve the lower channel access failure's probabilities compared to the URLLC UE in the multiple configuration scheme with the same number of the UE. When the number of the URLLC UE is bigger than 5, the multiple configuration scheme has a better channel access performance. The use of each scheme depends on the number of the UE and the UE's priorities in the system.


\section {Conclusion}\label{VI}
This paper analyzes channel access process in an unlicensed controlled  environment when the FBE channel access mechanism is used. The analysis through a Markov chain shows the limit of FBE to support URLLC due to a latency constraint. To improve the performance of channel access in FBE for URLLC, we propose two schemes. The first scheme allows the transmitter to use multiple FFP configurations to sense a channel and transmit data after a successful CCA. By using multiple FFP configurations, the transmitter has more chance to access to the channel in the URLLC latency budget. The second scheme configures FFPs of the transmitters in a system based on their priorities so that a high priority transmitter's transmission is not blocked by a lower priority transmitter's transmission. Therefore, by using one of two proposed schemes, the URLLC transmitter has a smaller channel blocking probability. Simulations have verified the analysis and shown the benefits of two proposed schemes compared to the current FBE schemes.

\end{document}